\documentclass[aps,pra,twocolumn,superscriptaddress]{revtex4}
\usepackage{amsfonts,amssymb,graphicx}
\usepackage[usenames,dvipsnames]{xcolor}
\usepackage{pdfpages}
\usepackage{footmisc}
\usepackage[normalem]{ulem}
\usepackage{color}

\begin{document}

\title{Ground-state and dynamical properties of a spin-$S$ Heisenberg star}
\author{Jiaxiu Li}
\affiliation{Center for Quantum Technology Research, School of Physics, Beijing Institute of Technology, Beijing 100081, China}
\affiliation{Key Laboratory of Advanced Optoelectronic Quantum Architecture and Measurements (MOE), School of Physics, Beijing Institute of Technology, Beijing 100081, China}
\author{Ning Wu}
\email{wunwyz@gmail.com}
\affiliation{Center for Quantum Technology Research, School of Physics, Beijing Institute of Technology, Beijing 100081, China}
\affiliation{Key Laboratory of Advanced Optoelectronic Quantum Architecture and Measurements (MOE), School of Physics, Beijing Institute of Technology, Beijing 100081, China}

\begin{abstract}
\par We generalize the Heisenberg star consisting of a spin-1/2 central spin and a homogeneously coupled spin bath modeled by the XXX ring [Richter J and Voigt A 1994 \emph{J. Phys. A: Math. Gen.} \textbf{27} 1139-1149] to the case of arbitrary central-spin size $S<N/2$, where $N$ is the number of bath spins. We describe how to block-diagonalize the model based on the Bethe ansatz solution of the XXX ring, with the dimension of each block Hamiltonian $\leq 2S+1$. We obtain all the eigenenergies and explicit expressions of the sub-ground states in each $l$-subspace with $l$ being the total angular momentum of the bath. Both the eigenenergies and the sub-ground states have distinct structures depending whether $S\leq l$ or $l<S$. The absolute ground-state energy and the corresponding  $l$ as functions of the intrabath coupling are numerically calculated for $N=16$ and $S=1,2,\cdots,7$ and their behaviors are quantitatively explained in the weak and strong intrabath coupling limits. We then study the dynamics of the antiferromagnetic order within an XXX bath prepared in the N\'eel state. Effects of the initial state of the central spin, the value of $S$, and the system-bath coupling strength on the staggered magnetization dynamics are investigated. By including a Zeeman term for the central spin and the anisotropy in the intrabath coupling, we also study the polarization dynamics of the central spin for a bath prepared in the spin coherent state. Under the resonant condition and at the isotropic point of the bath, the polarization dynamics for $S>1/2$ exhibits collapse-revival behaviors with fine structures. However, the collapse-revival phenomena is found to be fragile with respect to anisotropy of the intrabath coupling.
\end{abstract}

\maketitle

\section{Introduction}
\par Quantum spin systems are important physical systems that can exhibit many-body effects and strong correlations. They are ubiquitous in quantum magnetism, statistical physics, and more recently, quantum information and quantum simulations. It is generally challenging to theoretically study many-body spin systems due to the exponential growth of the dimension of the relevant Hilbert space with the system size. In this context, exactly soluble spin models play an important role in understanding the ground-state and dynamical properties of general large-scale spin systems.
\par Two important classes of soluble spin models are spin chains~\cite{Takahashi} and central spin models~\cite{Gaudin,RMP2004}, which can be solved by using free-fermion techniques or the Bethe ansatz. Currently, typical quantum spin chains such as the quantum Ising model and the XXZ chain have been realized on different experimental platforms~\cite{Trap,Ketterle2020} and continue to attract the attention of theorists~\cite{Balents2020,Suzuki}. Central spin models are highly relevant to solid-state setups that are promising candidates for performing quantum information processing, including electrons trapped in quantum dots~\cite{Loss2002} and nitrogen vacancy centers in diamond~\cite{phyrep}, etc.
\par In an early theoretical work, Richter and Voigt proposed a spin model that combines the above two types of soluble models, i.e., a spin-1/2 central spin model and an antiferromagnetic XXX periodic chain~\cite{RV}, with the intention of investigating the effect of central-spin induced frustration on the ground-state properties of the latter. Such a composite spin system, named as a Heisenberg star, is originally considered as an antiferromagnetic chain with a perturbation and has several conserved quantities that ensure the solvability of the model. Alternatively, the Heisenberg star can also be viewed as a central spin system in the presence of nearest-neighbor intrabath interactions. Recently, the solvability and real-time dynamics of higher-spin central spin models with/without intrabath interaction are studied and a richer variety of physical properties are observed compared to the spin-1/2 counterpart~\cite{Wu2020,Wu2022}.
\par In this work, we extend the spin-1/2 Heisenberg star to the case of a higher central spin of size $S\leq N/2$, where $N$ is the number of sites in the XXX ring. Following Ref.~\cite{Yang2020}, in which the coherence dynamics of a spin-1/2 Heisenberg star in the presence of an external magnetic was studied, we first describe how to block-diagonalize a spin-$S$ Heisenberg star based on the Bethe ansatz solutions of the XXX ring. The dimensions of the resultant block Hamiltonians are at most $2S+1$. With the help of the conserved quantities of the system, we then obtain all the eigenenergies of the model in terms of the quantum numbers $l$ and $j$, where $l$ and $j$ are the total angular momenta of the XXX bath and the whole system, respectively. Similar to the case of $S=1/2$, the sub-ground state energy in the $l$-subspace depends only on $l$. However, we find that these sub-ground state energies have different structures depending whether $S\leq l$ or $l<S$. Based on these results, we numerically calculate the absolute ground state energy and the corresponding bath angular momentum as functions of the intrabath coupling for an XXX bath of $N=16$ sites. The dependence of these quantities on varying $S$ is analytically analyzed in the weak intrabath coupling. We derive closed-form expressions of the $2j+1$ degenerate sub-ground states in the $l$-subspace, which can be written as a sum of tensor products of the central-spin state and the degenerate sub-ground states of the XXX bath, with the coefficients being determined analytically.
\par We are also interested in the real-time dynamics of the Heisenberg star. As observed in Ref.~\cite{Yang2020}, for a star prepared in a pure state the dynamics of any observable belonging to the central spin does not depend on the intrabath coupling, and hence is equivalent to the result for a noninteracting bath. Rather than focusing on the central spin dynamics, we study the staggered magnetization dynamics within the XXX bath when it is prepared in the N\'eel state. This is motivated by a theoretical investigation of the relaxation of antiferromagnetic order in a spin-1/2 XXZ chain following a quantum quench~\cite{PRL2009}. We reveal the influence of central-spin size, the central-spin initial state, and the system-bath coupling strength on the staggered magnetization dynamics. Some of the observed dynamical behaviors are consistent with those obtained in an inhomogeneous Heisenberg star~\cite{Wu2022}. For example, increasing the size of the central spin and adopting a superposed central-spin initial state can both accelerate the initial decay of the antiferromagnetic order, while these effects becomes less prominent in the strong intrabath coupling regime. It is intriguing that although the central spin dynamics is independent of the intrabath coupling, the magnetic order within the bath exhibits rich and robust dynamical behaviors even for homogeneous system-bath coupling.
\par We finally study the central-spin polarization dynamics for a slighted modified spin-$S$ Heisenberg star in the presence of an external magnetic field and with anisotropic intrabath coupling. Following Refs.~\cite{Dooley2013,Guan2019,PRA2020}, we choose the  spin coherent state as the bath initial state. We demonstrate that at the isotropic point of the bath the central spin dynamics from the spin coherent state is the same as that for a noninteracting bath. For an XXX bath with $S=1/2$, we recover the prior results presented in Ref.~\cite{Guan2019}. For $S>1/2$,  we find that the polarization dynamics exhibits collapse-revival behaviors with fine structures under the resonant condition. However, the collapse-revival phenomena are destroyed once the anisotropic intrabath coupling is introduced.
\par The rest of the paper is organized as follows. In Sec.~\ref{SecII}, we introduce the spin-$S$ Heisenberg star and its conserved quantities. In Sec.~\ref{SecIII}, we study the eigenenergies of the model in detail and obtain expressions of the sub-ground state energies in each $l$-subspace. In Sec.~\ref{SecIV}, we describe the block diagonalization procedure using the Bethe ansatz solution of the XXX bath and derive explicit expressions for the degenerate sub-ground states in each $l$-subspace. In Sec.~\ref{SecV}, we study in detail the dynamics of the Heisenberg star for baths prepared in the N\'eel state and the spin coherent state. Conclusions are drawn in Sec.~\ref{SecVI}.
\section{Model and conserved quantities}\label{SecII}
\par
\begin{figure}
\includegraphics[width=.52\textwidth]{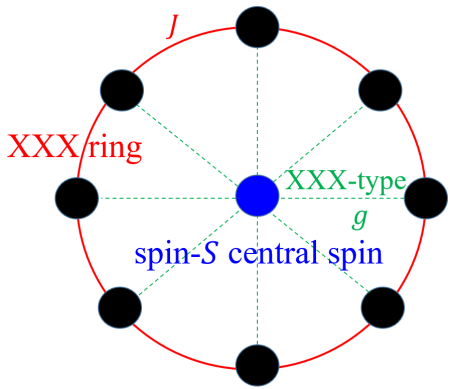}
\caption{A spin-$S$ Heisenberg star consists of a central spin of size $S$ and a homogeneously coupled XXX ring. The system-bath (intrabath) coupling is of XXX-type with strength $g$ ($J$).}
\label{Fig1}
\end{figure}

 The Heisenberg star was first introduced in Ref.~\cite{RV} and is described by the Hamiltonian (see Fig.~\ref{Fig1})
\begin{eqnarray}\label{Hstar}
H&=&H_{\rm{B}}+H_{\rm SB},\nonumber\\
H_{\rm B}&=& J\sum^N_{n=1}\vec{S}_n\cdot\vec{S}_{n+1},\nonumber\\
H_{\rm SB}&=&g\vec{S}\cdot \sum^N_{n=1}\vec{S}_n.
\end{eqnarray}
Here, $H_{\rm B}$ describes a spin-1/2 periodic Heisenberg XXX spin chain with antiferromagnetic nearest-neighbor coupling strength $J>0$. The interaction between the central spin $\vec{S}$ with size $S$ and the XXX spin bath is of isotropic Heisenberg type and the coupling strength is measured by $g>0$. The static properties of $H$ for $S=1/2$ were studied in detail in Ref.~\cite{RV}, but here we allow for arbitrary values of $S$ with $S<N/2$.
\par We define the total spin of the whole system as
\begin{eqnarray}
\vec{\mathcal{J}}=\vec{S}+\vec{L},
\end{eqnarray}
where
\begin{eqnarray}
\vec{L}=\sum^N_{n=1}\vec{S}_n
\end{eqnarray}
is the total spin of the XXX bath. We can rewrite $H_{\rm SB}$ in terms of $\vec{S}$ and $\vec{L}$ as
\begin{eqnarray}
H_{\rm SB}&=&g\vec{S}\cdot\vec{L}.
\end{eqnarray}
\par It is easy to check the following commutation relations 
\begin{eqnarray}\label{HHB}
~[H,\vec{\mathcal{J}}^2]&=&0,~~[H,\mathcal{J}_z]=0,\nonumber\\
~[H,\vec{L}^2]&=&0,\nonumber\\
~[H,H_{\rm B}]&=&[H_{\rm SB},H_{\rm B}]=0.
\end{eqnarray}
\par The last relation indicates that we can diagonalize $H_{\rm B}$ and $H_{\rm SB}$ separately. Due to the high symmetry of the model, any eigenstate $|\psi_{E,j,m,l}\rangle$ can be labelled by four quantum numbers $E, j, m, l$, which belong to the conserved quantities $H$, $\vec{\mathcal{J}}^2$, $\mathcal{J}_z$, $\vec{L}^2$, respectively. For simplicity, we assume that $N$ is an even integer. In the discussion of static properties of the spin-$S$ Heisenberg star (Sec.~\ref{SecII}), we further assume that $S$ is an integer (the case of half-odd-integer $S$ can be similarly analyzed). In order to obtain universal size-independent results, we use the collective coupling $\tilde{g}\equiv g\sqrt{N}$ as an overall energy scale throughout this work.
\par From the relation $\vec{S}\cdot\vec{L}=(\vec{\mathcal{J}}^2-\vec{S}^2-\vec{L}^2)/2$, we can further rewrite $H$ as
\begin{eqnarray}\label{Hjsl}
H=JH_{\rm b}+\frac{g}{2}(\vec{\mathcal{J}}^2-\vec{S}^2-\vec{L}^2),
\end{eqnarray}
where $H_{\rm b}=\sum^N_{n=1}\vec{S}_n\cdot\vec{S}_{n+1}$.
\section{Eigenenergies}\label{SecIII}
\par All the eigenenergies of the spin-$S$ Heisenberg star can be determined from Eq.~(\ref{Hjsl}) once the spectrum of the bath Hamiltonian $H_{\rm b}$ is solved. We first review the eigenenergy structure of an isolated XXX ring, based on which we construct all the eigenenergies of the Heisenberg star.
\subsection{Energy levels of an isolated XXX ring}
\par For completeness, let us first review some known results about the pure XXX ring described by $H_{\rm b}$. It is well known that the eigenenergies and eigenstates of the XXZ chain can be solved by using the coordinate Bethe ansatz within individual sectors possessing fixed magnetization~\cite{Takahashi}. For the isotropic XXX chain, the total angular momentum $\vec{L}$ is further conserved, yielding simpler state structures and increasing degrees of degeneracy of the energy levels.
\par For even $N$, the addition of the $N$ spins-1/2 in the ring results in $N/2+1$ total angular momenta $l=0,1,\cdots,N/2$, where a fixed $l$ appears ($C^n_m=\frac{m!}{n!(m-n)!}$ is the binomial coefficient and vanishes for $n>m$)
\begin{eqnarray}
d_{N,l}=C^{l+N/2}_{N}-C^{l+1+N/2}_{N}
\end{eqnarray}
times~\cite{Dicke}. Since $[H_{\rm b},\vec{L}^2]=[H_{\rm b},L_z]=0$, any eigensate of $H_{\rm b}$ can be written as $|\phi_{E^{(\alpha_l)}_{\rm b}(l),l,l_m}\rangle$, where $E^{(\alpha_l)}_{\rm b}(l)$ is the corresponding eigenenergy with the superscript $\alpha_l=1,2,\cdots,d_{N,l}$ distinguishing the energy levels having the same value of $l$, and $l_m$ is the eigenvalue of $L_z$. Note that $E^{(\alpha_l)}_{\rm b}(l)$ does not depend on $l_m$ and is $(2l+1)$-fold degenerate, with the corresponding degenerate eigenstates $\{|\phi_{E^{(\alpha_l)}_{\rm b},l,-l}\rangle,|\phi_{E^{(\alpha_l)}_{\rm b},l,-l+1}\rangle,\cdots,|\phi_{E^{(\alpha_l)}_{\rm b},l, l}\rangle\}$. It is easy to check that
\begin{eqnarray}
\sum^{N/2}_{l=0}(2l+1)d_{N,l}=2^N,
\end{eqnarray}
leading to a consistency.
\par If we assume that $E^{(1)}_{\rm b}(l)\leq E^{(2)}_{\rm b}(l)\leq\cdots\leq E^{(d_{N,l})}_{\rm b}(l)$, then the Lieb-Mattis-Marshall theorem~\cite{Lieb,Marshall} tells us that the sub-ground state energy $E^{(1)}_{\rm b}(l)$ in every $l$-subspace satisfies
\begin{eqnarray}
E^{(1)}_{\rm b}(l)<E^{(1)}_{\rm b}(l+1).
\end{eqnarray}
Moreover, the lowest-energy state for fixed $l_m$ is just $|\phi_{E^{(1)}_{\rm b},|l_m|,l_m}\rangle$, which is nondegenerate in this $l_m$-subspace and possesses energy $E^{(1)}_{\rm b}(|l_m|)$. In other words, $E^{(1)}_{\rm b}(l)$ is the lowest energy level in magnetization sector with $l_m=l$. A direct consequence of these results is that the global ground state of $H_{\rm b}$ is a unique singlet $|\phi_{E^{(1)}_{\rm b},0,0}\rangle$.
\subsection{Eigenenergies of the spin-$S$ Heisenberg star}
\par Let us now turn back to the spin-$S$ Heisenberg star. Depending on whether $S\leq l$ or $S>l$, the total angular momentum $j$ of the star is accordingly determined by:
\par ~
\par 1) For fixed $l$ satisfying $S\leq l\leq \frac{N}{2}$, the addition of $l$ and $S$ gives the following $2S+1$ different values of $j$
\begin{eqnarray}
j=l+s,~~s=-S, -S+1,\cdots, S.
\end{eqnarray}
According to Eq.~(\ref{Hjsl}), for a given $l$, the eigenenergy for a fixed $s$ (and hence for a fixed $j=l+s$) is
\begin{eqnarray}\label{Els1}
E^{(\alpha_l)}(l,s)=JE^{(\alpha_l)}_{\rm b}(l)+\frac{g}{2}[s^2+s(2l+1)-S(S+1)].\nonumber\\
\end{eqnarray}
The energy level $E^{(\alpha_l)}(l,s)$ is $(2j+1)$-fold degenerate since the $2j+1$ states $|\psi_{E^{(\alpha_l)}(l,s),j,m,l}\rangle ~(m=-j,-j+1,\cdots,j)$ possess the same energy and are connected by the raising or lowering operator $\mathcal{J}_\pm$. We are interested in the lowest eigenenergy for a fixed $l$, i.e., the sub-ground state energy in the $l$-subspace. By noting that $s^2+s(2l+1)$ is an increasing function of $s$ for $s>-(l+1/2)$, the second term in Eq.~(\ref{Els1}) is minimized for $s=-S$. Thus, the sub-ground state energy in the $l$-subspace with $S\leq l\leq\frac{N}{2}$ is
\begin{eqnarray}\label{EGl1}
E^{(\mathrm{gs})}(l)\equiv E^{(1)}(l,-S)=JE^{(1)}_{\rm b}(l)-gS(l+1),
\end{eqnarray}
which is $[2(l-S)+1]$-fold degenerate with the corresponding eigenstates $\{|\psi_{E^{(\mathrm{gs})}(l),l-S,m,l}\rangle\}$, $-(l-S)\leq m\leq l-S$. Since $JE^{(1)}_{\rm b}(l)$ is an increasing function of $l$ for $J>0$ and $-gS(l+1)$ is a decreasing function of $l$ for $g>0$, there exists a competition between the two terms in $E^{(\mathrm{gs})}(l)$ and there must be some $l=l_>$ that minimizes $E^{(\mathrm{gs})}(l)$.
\par ~
\par 2) For fixed $l$ satisfying $0\leq l<S$, $j$ can take values
\begin{eqnarray}
j=S+s,~~s=-l,-l+1,\cdots,l.
\end{eqnarray}
The eigenenergy for fixed $l$ and $j$ is
\begin{eqnarray}\label{Els2}
E^{(\alpha_l)}(l,s)=JE^{(\alpha_l)}_{\rm b}(l)+\frac{g}{2}[s^2+s(2S+1)-l(l+1)].\nonumber\\
\end{eqnarray}
Similarly, the sub-ground state in the $l$-subspace with $0\leq l<S$ is achieved for $s=-l$:
\begin{eqnarray}\label{EGl2}
E^{(\mathrm{gs})}(l)\equiv E^{(1)}(l,-l)=JE^{(1)}_{\rm b}(l)-gl(S+1),
\end{eqnarray}
which is $[2(S-l)+1]$-fold degenerate. There exists a certain $l=l_<$ that minimizes $E^{(\mathrm{gs})}(l)$ for $0\leq l<S$.
\par ~
\par Once $E^{(\mathrm{gs})}(l_<)$ and $E^{(\mathrm{gs})}(l_>)$ are obtained, the absolute ground state energy of $H$ is simply
\begin{eqnarray}
E^{(G)}(l^{(G)})=\min\{E^{(\mathrm{gs})}(l_<),E^{(\mathrm{gs})}(l_>)\},
\end{eqnarray}
where $l^{(G)}$ is the total angular momentum of the bath in the global ground state.
\begin{figure}
\includegraphics[width=.53\textwidth]{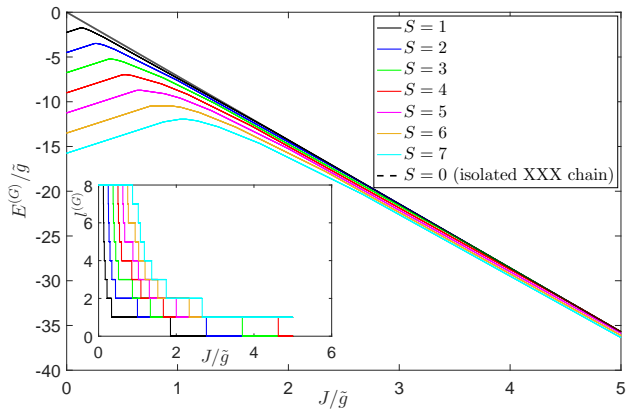}
\caption{Main panel: The ground-state energy $E^{(G)}/\tilde{g}$ as a function of $J/\tilde{g}$ for a spin-$S$ Heisenberg star with $N=16$ bath spins. Results for $S=1,2,\cdots,7$ are shown. The dashed black line represents the result for a pure XXX chain or a Heisenberg star with $S=0$. Inset: The evolution of total angular momentum of the bath, $l^{(G)}$, as $J/\tilde{g}$ increases.}
\label{Fig2}
\end{figure}
\par The main panel of Fig.~\ref{Fig2} shows the ground-state energy $E^{(G)}(l^{(G)})$ as a function of $J/\tilde{g}$ for $N=16$ and $S=1,2,\cdots,7$. We observe that:
\par i) For fixed $J/\tilde{g}$, the ground-state energy $E^{(G)}$ decreases with increasing $S$. In the large $J/\tilde{g}$ limit, $E^{(G)}$ converges to the result of $S=0$ (or of a pure XXX chain) for different values of $S$.
\par ii) For fixed $S$, $E^{(G)}$ is a nonmonotonic function of $J/\tilde{g}$, indicating that there might exist level crossings at certain values of $J/\tilde{g}$.
\par iii) In the small $J/\tilde{g}$ limit, $E^{(G)}$ increases linearly with increasing $J/\tilde{g}$ and the energy difference for adjacent $S$'s is a constant.
\par The above behaviors of $E^{(G)}$ can be understood by inspecting Eqs.~(\ref{EGl1}) and (\ref{EGl2}). In the large $J/\tilde{g}$ limit, the system-bath coupling can be viewed as a perturbation for the XXX ring and the first terms in Eqs.~(\ref{EGl1}) and (\ref{EGl2}) are dominated, which explains the convergence of $E^{(G)}$ to the result for $S=0$.
\par Since the ground-state energy $E^{(G)}(l^{(G)})$ depends only on the bath angular momentum $l^{(G)}$, it is expected that the nonmonotonic behavior of $E^{(G)}(l^{(G)})$ and the associated level crossings are caused by the sudden change of $l^{(G)}$. To this end, we plot in the inset of Fig.~\ref{Fig2} the evolution of $l^{(G)}$ with varying $J/\tilde{g}$. It can be seen that $l^{(G)}$ shows plateaus that decrease from $l^{(G)}=N/2$ to $l^{(G)}=0$ as $J/\tilde{g}$ increases. Actually, in the small $J/\tilde{g}$ limit, the second terms in the ground-state energy $E^{(\mathrm{gs})}(l)$ dominate, resulting in $E^{(\mathrm{gs})}(l)\approx -gS(l+1)$ for $S\leq l\leq N/2$ and $E^{(\mathrm{gs})}(l)\approx -gl(S+1)$ for $0\leq l<S$. For fixed $S<N/2$, it is apparent that $l^{(G)}$ tends to take its largest possible value, i.e., $l^{(G)}=N/2$, as can be seen from the inset of Fig~\ref{Fig2}. We also observe that the transition point from $l^{(G)}=N/2$ to $l^{(G)}=N/2-1$ increases with increasing $S$. To understand this phenomenon as well as observation (iii), we will look at the small $J/\tilde{g}$ limit in detail, where some analytical results for $E^{(G)}$ are available.
\subsection{Analytical results in the small $J/\tilde{g}$ limit}
\par For $l^{(G)}=N/2$ and $N/2-1$, the sub-ground state energy $E^{(1)}_{\rm b}(l^{(G)})$ of the XXX chain admits analytical expressions. To find out the transition point from $l^{(G)}=N/2$ to $l^{(G)}=N/2-1$, we first study the case of $l^{(G)}=N/2>S$. From Eq.~(\ref{EGl1}), we have
\begin{eqnarray}
\frac{E^{(G)}(\frac{N}{2})}{\tilde{g}}&=&\frac{J}{\tilde{g}}E^{(1)}_{\rm b}\left(\frac{N}{2}\right) -\frac{g}{\tilde{g}}S\left(\frac{N}{2}+1\right)\nonumber\\
&=&\frac{N}{4}\frac{J}{\tilde{g}}-\frac{S}{\sqrt{N}}\left(\frac{N}{2}+1\right),
\end{eqnarray}
where we used $E^{(1)}_{\rm b}\left(\frac{N}{2}\right)=N/4$ for the fully polarized state. This explains the initial linear increase of $E^{(G)}(l^{(G)})/\tilde{g}$ before the first transition occurs. The slope is $N/4$ for all $S$ and the energy difference for adjacent $S$'s is a constant $\frac{1}{\sqrt{N}}(N/2+1)$ in this linear region.
\par For $l^{(G)}=N/2-1$, the condition $S\leq l^{(G)}$ is still satisfied. We thus have $E^{(G)}(N/2-1)/\tilde{g}=JE^{(1)}_{\rm b}(N/2-1)/\tilde{g}-gS\frac{N}{2}/\tilde{g}$. According to the Lieb-Mattis-Marshall theorem, $E^{(G)}(N/2-1)$ is the eigenenergy of the lowest single-magnon state. It is known that the single-magnon dispersion for the XXX chain $H_{\rm b}$ is~\cite{Wumagnon}
\begin{eqnarray}
\mathcal{E}_1(k)=\frac{N}{4}-(1-\cos k), ~~\mathrm{e}^{\mathrm{i} kN}=1,
\end{eqnarray}
which gives $E^{(1)}_{\rm b}(\frac{N}{2}-1)=\mathcal{E}_1(k=0)=\frac{N}{4}-2$, and hence
\begin{eqnarray}
\frac{E^{(G)}(\frac{N}{2}-1)}{\tilde{g}}= \left(\frac{N}{4}-2\right)\frac{J}{\tilde{g}}-\frac{1}{2}S\sqrt{N}.
\end{eqnarray}
We see that $E^{(G)}(N/2-1)/\tilde{g}$ is also a linearly increasing function of $J/\tilde{g}$ before the transition $l^{(G)}=N/2-1\to N/2-2$ occurs, but the slope is reduced to $N/4-2$ compared to the case of $l^{(G)}=N/2$.
The transition point for $l^{(G)}=N/2\to N/2-1$ is determined by
\begin{eqnarray}
\frac{E^{(G)}(\frac{N}{2})}{\tilde{g}}=\frac{E^{(G)}(\frac{N}{2}-1)}{\tilde{g}},
\end{eqnarray}
yielding the transition coupling strength
\begin{eqnarray}
\frac{J}{\tilde{g}}=\frac{S}{2\sqrt{N}},
\end{eqnarray}
which is actually a linear function of $S$ (inset of Fig.~\ref{Fig2}, the top horizontal lines).
\section{Eigenstates}\label{SecIV}
\par Let us now turn to study the eigenstates of the spin-$S$ Heisenberg star. We first briefly describe how to obtain all the eigenstates of the Heisenberg star with the help the Bethe ansatz solution of the XXX chain, which divides the full Hilbert space into invariant subspaces of at most $2S+1$ dimensions. We then focus on the sub-ground states within the sector with fixed $l$ and derive closed-from expressions of these states in terms of the sub-ground states of the XXX chain in the $l$-sector.
\subsection{General eigenstates: the Bethe ansatz method}
\par We closely follow Refs.~\cite{Yang2020,BAbook} to construct invariant subspaces of $H$ based on the Bethe ansatz solution of the XXX bath $H_{\rm B}$. The Bethe states of $H_{\rm B}$ with $M\leq N/2$ spin flips are of the form
\begin{eqnarray}\label{BAs}
|\lambda_1,\cdots,\lambda_M\rangle=B(\lambda_1)\cdots B(\lambda_M)|F\rangle,
\end{eqnarray}
where $\{\lambda_j\}$ are the Bethe roots determined by the Bethe ansatz equations, $B(\lambda_i)$ is the spin-flipping operator appearing in the so-called monodromy matrix~\cite{BAbook}, and $|F\rangle=|\uparrow\cdots\uparrow\rangle$ is the fully polarized reference state.
\par It is known that the Bethe state given by Eq.~(\ref{BAs}) is the highest weight state of the $su(2)$ Lie algebra generated by the bath operators $(L_\pm,L_z)$. By successively applying the lowering operator $L_-$ to the Bethe states, we can obtain the $(N-2M+1)$-fold degenerate manifold corresponding to the eigenenergy $E_{\rm B}(\lambda_1,\cdots,\lambda_M)$, which satisfies the Schr\"odinger equation $H_{\rm B}|\lambda_1,\cdots,\lambda_M\rangle=E_{\rm B}(\lambda_1,\cdots,\lambda_M)|\lambda_1,\cdots,\lambda_M\rangle$. Explicitly, we define~\cite{Yang2020}
\begin{eqnarray}\label{BAdeg}
|\lambda_1,\cdots,\lambda_M;n\rangle=C_{M,n}(L_-)^{\mathcal{M}-n}|\lambda_1,\cdots,\lambda_M\rangle,
\end{eqnarray}
where $\mathcal{M}=N/2-M$ is the magnetization of the Bethe state $|\lambda_1,\cdots,\lambda_M\rangle$, $n=\mathcal{M},\mathcal{M}-1,\cdots,-\mathcal{M}$, and $C_{M,n}$ is a suitable normalization coefficient. The degenerate states given by Eq.~(\ref{BAdeg}) satisfy the following relations,
\begin{eqnarray}
&&L_z|\lambda_1,\cdots,\lambda_M;n\rangle=n|\lambda_1,\cdots,\lambda_M;n\rangle,\nonumber\\
&&L_{\pm}|\lambda_1,\cdots,\lambda_M;n\rangle=\sqrt{(\mathcal{M}\mp n)(\mathcal{M}\pm n+1)}\nonumber\\
&&\times|\lambda_1,\cdots,\lambda_M;n\pm 1\rangle.
\end{eqnarray}
It is obvious that $L_\pm |\lambda_1,\cdots,\lambda_M;\pm\mathcal{M}\rangle=0$.
\par Having the above eigenstates of the XXX chain in hand, the base states for the spin-$S$ star are
\begin{eqnarray}\label{base}
\{|S_m\rangle |\lambda_1,\cdots,\lambda_M;n\rangle\},\nonumber
\end{eqnarray}
where $S_m=S,S-1,\cdots,-S$, $M=0,1,\cdots,N/2$, and $n=\mathcal{M},\mathcal{M}-1,\cdots,-\mathcal{M}$. By rewriting the Hamiltonian as $H=H_{\rm B}+\frac{1}{2}g(S_+L_-+S_-L_+)+gS_zL_z$, it can be easily seen that the two states $|S\rangle |\lambda_1,\cdots,\lambda_M;\mathcal{M}\rangle$ and $|-S\rangle |\lambda_1,\cdots,\lambda_M;-\mathcal{M}\rangle$ are simple eigenstates of $H$ with the same eigenenergy $E_{\rm B}(\lambda_1,\cdots,\lambda_M)+gS\mathcal{M}$. For $S_m+n\neq \pm \mathcal{M}$, we apply $H$ to the state $|S_m\rangle |\lambda_1,\cdots,\lambda_M;n\rangle$ to get
\begin{eqnarray}
&&H|S_m\rangle |\lambda_1,\cdots,\lambda_M;n\rangle=(E_{\rm B}+gn)|S_m\rangle |\lambda_1,\cdots,\lambda_M;n\rangle\nonumber\\
&&+\frac{g}{2}\sqrt{(S-S_m)(S+S_m+1)(\mathcal{M}+ n)(\mathcal{M}- n+1)}\nonumber\\
&&\times|S_m+1\rangle |\lambda_1,\cdots,\lambda_M;n-1\rangle\nonumber\\
&&+\frac{g}{2}\sqrt{(S+S_m)(S-S_m+1)(\mathcal{M}-n)(\mathcal{M}+n+1)}\nonumber\\
&&\times|S_m-1\rangle |\lambda_1,\cdots,\lambda_M;n+1\rangle.
\end{eqnarray}
Unlike the case of $S=1/2$ where $|1/2\rangle |\lambda_1,\cdots,\lambda_M;n\rangle$ and $|-1/2\rangle |\lambda_1,\cdots,\lambda_M;n+1\rangle$ already form a closed subspace~\cite{Yang2020}, for a general $S\leq N/2$ we need to further apply $H$ to the newly generated states $|S_m+1\rangle |\lambda_1,\cdots,\lambda_M;n-1\rangle$ and $|S_m-1\rangle |\lambda_1,\cdots,\lambda_M;n+1\rangle$ to obtain a multi-dimensional invariant subspace.
\par For $S\leq N/2$, these invariant subspaces can be classified into three categories:
\par I) For $-S-N/2\leq m\leq S-N/2$ with $m=S_m+n$ the total magnetization of the star, the configurations of $(S_m,n)$ that conserve $m$ are $(-S,m+S),\cdots,(m+N/2,-N/2)$. The dimension of the corresponding invariant subspace is therefore $m+N/2+S+1\leq 2S+1$.
\par II) For $S-N/2+1\leq m\leq -S+N/2-1$, the configurations of $(S_m,n)$ that conserve the total magnetization are $(-S,m+S),\cdots,(S,m-S)$. The dimension of the corresponding invariant subspace is $2S+1$.
\par III) For $-S+N/2\leq m\leq S+N/2$, the configurations of $(S_m,n)$ that conserve the total magnetization are $(m-N/2,N/2),\cdots,(S,m-S)$. The dimension of the corresponding invariant subspace is $S-m+N/2+1\leq 2S+1$.
\par In this way, the whole Hilbert space of the spin-$S$ Heisenberg star is divided into invariant subspaces whose dimensions are at most $2S+1$. In principle, we can numerically diagonalize the block Hamiltonians to obtain all the eigenstates and eigenenergies of the system.
\par The Bethe ansatz method presented above is also applicable when a Zeeman term $\omega S_z$ of the central spin is included~\cite{Yang2020}. In this case, the total angular momentum $\vec{\mathcal{J}}$ is no longer conserved since $[\vec{\mathcal{J}}^2,S_z]=2[\vec{S}\cdot\vec{L},S_z]\neq 0$, so that the state $|\psi_{E,j,m,l}\rangle$ is not well-defined. However, if the Heisenberg star $H$ given by Eq.~(\ref{Hstar})  (in the absence of the Zeeman term) is mainly concerned, the states $\{|\psi_{E,j,m,l}\rangle\}$ provide a more convenient form to analyze the eigenstate structure of $H$.
\par Corresponding to the two types of eigenenergies given by Eqs.~(\ref{Els1}) and (\ref{Els2}), the number of eigenstates $\{|\psi_{E^{(\alpha_l)},j,m,l}\rangle\}$ can be counted as
\begin{eqnarray}
\mathcal{N}&=&\sum^{\frac{N}{2}}_{l=S}d_{N,l}\sum^S_{s=-S}[2(l+s)+1]\nonumber\\
&&+\sum^{S-1}_{l=0}d_{N,l}\sum^l_{s=-l}[2(S+s)+1].
\end{eqnarray}
It can be checked that $\mathcal{N}$ is identical to the total dimension of the Hilbert space $(2S+1)2^N$. Below we focus on the sub-ground states for fixed $l$'s. As we will see, these sub-ground states admit closed-form expressions in terms of the sub-ground states of the pure XXX ring.
\subsection{Sub-ground states for $S\leq l\leq\frac{N}{2}$}
\par For fixed $l\geq S$, the sub-ground states $\{|\psi_{E^{(\mathrm{gs})}(l),l-S,m,l}\rangle\}$ have total spin $j=l-S$, and satisfy
\begin{eqnarray}\label{SchHbG}
H|\psi_{E^{(\mathrm{gs})}(l),l-S,m,l}\rangle=E^{(\mathrm{gs})}(l)|\psi_{E^{(\mathrm{gs})}(l),l-S,m,l}\rangle,
\end{eqnarray}
where $-(l-S)\leq m\leq l-S$ and $E^{(\mathrm{gs})}(l)=JE^{(1)}_{\rm b}(l)-gl(S+1)$. For fixed $m$, we have $-l\leq m-S_m\leq l$ for all $-S\leq S_m\leq S$, so that $|\psi_{E^{(\mathrm{gs})}(l),l-S,m,l}\rangle$ is of the form
\begin{eqnarray}
|\psi_{E^{(\mathrm{gs})}(l),l-S,m,l}\rangle=\sum^S_{S_m=-S}A_{S_m}|S_m\rangle|\phi_{E^{(1)}_{\rm b},l,m-S_m}\rangle,
\end{eqnarray}
where $|S_m\rangle$ is the eigenstate of $S_z$ with eigenvalue $S_m$, $|\phi_{E^{(1)}_{\rm b},l,m-S_m}\rangle$ is the lowest eigenstates of $H_{\rm b}$ for fixed $l$ and $l_m$ with energy $E^{(1)}_{\rm b}(l)$, and the $A$'s are coefficients to be determined by Eq.~(\ref{SchHbG}).
\par After a tedious but straightforward calculation, we arrive at the following \emph{unnormalized} sub-ground state (see the Appendix for the derivation)
\begin{eqnarray} \label{Asmfinal}
&&|\psi_{E^{(\mathrm{gs})}(l),l-S,m,l}\rangle=\sum^S_{S_m=-S}(-1)^{S-S_m}\sqrt{C^{S+S_m}_{2S}}\nonumber\\
&&\times\sqrt{\frac{(l+m-S_m)!(l-m+S_m)!}{(l+m-S)!(l-m+S)!}}|S_m\rangle|\phi_{E^{(1)}_{\rm b},l,m-S_m}\rangle.\nonumber\\
\end{eqnarray}
In particular, the highest-weight state $|\psi_{E^{(\mathrm{gs})}(l),l-S,l-S,l}\rangle$ can be normalized as (see Appendix)
\begin{eqnarray} \label{AsmfinalHW}
&&|\psi_{E^{(\mathrm{gs})}(l),l-S,l-S,l}\rangle= \sqrt{\frac{(2l-2S+1)(2S)!}{(2l+1)!}}\nonumber\\
&&\times\sum^S_{S_m=-S}(-1)^{S-S_m}  \sqrt{\frac{(2l-S-S_m)! }{ (S-S_m)! }}|S_m\rangle|\phi_{E^{(1)}_{\rm b},l,l-S-S_m}\rangle.\nonumber\\
\end{eqnarray}
Note that the sub-ground states do not depend on the coupling strengths $J$ and $g$ but are determined by the quantum number $l$.
\subsection{Sub-ground states for $0\leq l < S$}
\par For fixed $0\leq l<S$, the sub-ground states $\{|\psi_{E^{(\mathrm{gs})}(l),S-l,m,l}\rangle\}$ have total angular momentum $j=S-l$ and satisfy
\begin{eqnarray}\label{SchHbG1}
H|\psi_{E^{(\mathrm{gs})}(l),S-l,m,l}\rangle=E^{(\mathrm{gs})}(l)|\psi_{E^{(\mathrm{gs})}(l),S-l,m,l}\rangle,
\end{eqnarray}
where $-(S-l)\leq m\leq S-l$. For fixed $m$, we have $-S\leq m-l_m\leq S $ for all $-l\leq l_m\leq l$. The most general form of $|\psi_{E^{(\rm gs)}(l),S-l,m,l}\rangle$ is
\begin{eqnarray}
|\psi_{E^{(\mathrm{gs})}(l),S-l,m,l}\rangle=\sum^{l}_{l_m=-l}B_{l_m}|m-l_m\rangle|\phi_{E^{(1)}_{\rm b},l,l_m}\rangle.
\end{eqnarray}
It can be similarly shown that [by making the substitutions $S\to l,~l\to S,~S_m\to l_m$ in Eq.~(\ref{Asmfinal})]
\begin{eqnarray} \label{Blmfinal}
&&|\psi_{E^{(\mathrm{gs})}(l),S-l,m,l}\rangle=\sum^{l}_{l_m=-l}(-1)^{l-l_m}\sqrt{C^{l+l_m}_{2l}}\nonumber\\
&&\times\sqrt{\frac{(S+m-l_m)!(S-m+l_m)!}{(S+m-l)!(S-m+l)!}}|m-l_m\rangle|\phi_{E^{(1)}_{\rm b},l,l_m}\rangle.\nonumber\\
\end{eqnarray}
\section{Real-time dynamics}\label{SecV}
\par The spin-$S$ Heisenberg star given by Eq.~(\ref{Hstar}) is so special that it cannot generate any intrabath coupling-induced central-spin dynamics. Suppose $\chi_{\rm S}$ is an arbitrary observable belonging to the central spin, its time evolution from an initial state $|\psi(0)\rangle$ is given by
\begin{eqnarray}
\langle\chi_{\rm S}(t)\rangle=\langle\psi(0)|\mathrm{e}^{\mathrm{i}Ht}\chi_{\rm S}\mathrm{e}^{-\mathrm{i}Ht}|\psi(0)\rangle.
\end{eqnarray}
From $H=JH_{\rm b}+g\vec{S}\cdot\vec{L}$ and $[H_{\rm b},\chi_{\rm S}]=0$, we have $[H,\chi_{\rm S}]=g[\vec{S}\cdot\vec{L},\chi_{\rm S}]$, which is independent of the intrabath coupling $J$. By noting that $H_{\rm b}$ is rotationally invariant, i.e., $[H_{\rm b},L_i]=0$, we further have $[H,[H,\chi_{\rm S}]]=g^2[\vec{S}\cdot\vec{L},[\vec{S}\cdot\vec{L},\chi_{\rm S}]]$, $[H,[H,[H,\chi_{\rm S}]]]=g^3[\vec{S}\cdot\vec{L},[\vec{S}\cdot\vec{L},[\vec{S}\cdot\vec{L},\chi_{\rm S}]]]$,$\cdots$, giving $\langle\chi_{\rm S}(t)\rangle=\langle\psi(0)|\mathrm{e}^{\mathrm{i}H_{\rm SB}t}\chi_{\rm S}\mathrm{e}^{-\mathrm{i}H_{\rm SB}t}|\psi(0)\rangle$. In other words, the time evolution of any central-spin observable is independent of the intrabath coupling $J$, and hence recovers the result for a noninteracting bath.
\par It is easy to see that including any central-spin term (such as a Zeeman term or a single-ion anisotropy, etc.) does not change the foregoing property, as already been observed in the investigation of the central spin coherence from a pure state~\cite{Yang2020}. Thus, to obtain nontrivial dynamics of the central spin induced by the intrabath coupling, one has to either include the thermal effect~\cite{Yang2020} or to go beyond the homogeneous system-bath coupling or isotropic intrabath coupling~\cite{Wu2022}. In spite of these facts, the dynamics of any bath operator $\eta_{\rm B}$ depends on both $J$ and $g$ since generally $[H_{\rm b},\eta_{\rm B}]\neq 0$ and $[\vec{S}\cdot\vec{L},\eta_{\rm B}]\neq0$.
\par In this work, we first study the dynamics of the antiferromagnetic order in the XXX bath governed by the spin-$S$ Heisenberg star $H$, with the XXX bath prepared in a N\'eel state. We then study the central spin dynamics in a slighted generalized Heisenberg star with intrabath anisotropy. To be specific, in this case we choose the bath initial state as a spin coherent state.
\subsection{Dynamics of antiferromagnetic order in the spin-$S$ Heisenberg star}
\par The dynamics of antiferromagnetic order in an XXZ bath with inhomogeneous system-bath coupling and anisotropic intrabath coupling has been thoroughly studied in a related work by the authors~\cite{Wu2022}. Compare with the case of an isolated XXZ chain~\cite{PRL2009}, it is found that both the system-bath coupling and the size of the central spin have significant influence on the relaxation of the antiferromagntic order. We show that some of the dynamical behaviors of the antiferromagnetic order observed in Ref.~\cite{Wu2022} for inhomogeneous system-bath couplings are still robust in the homogeneous Heisenberg star described by $H$.
\par We assume that the star is initially prepared in a product state
\begin{eqnarray}\label{psi0SCS}
|\psi(0)\rangle&=&|\phi^{(\rm S)}\rangle\otimes|\mathrm{AF}\rangle,
\end{eqnarray}
where $|\phi^{(\rm S)}\rangle$ is the initial state of the central spin and the bath initial state is chosen as the N\'eel state $|\mathrm{AF}\rangle=|\downarrow\uparrow\cdots\downarrow\uparrow\rangle$. For arbitrary $S<N/2$, the dynamics of the spin-$S$ Heisenberg star $H$ is simulated by using an equations-of-motion method based on analytical expressions of spin-operator matrix elements for the XX chain~\cite{PRB2018}, see Ref.~\cite{Wu2022} for details of the method.
\par We consider two types of initial states for the central spin, i.e., the polarized state $|\phi^{(\rm S)}\rangle_1=|S\rangle$ and the equally weighted superposition state $|\phi^{(\rm S)}\rangle_2=\frac{1}{\sqrt{2S+1}}(|S\rangle+|S-1\rangle+\cdots+|-S\rangle)$. We are interested in the time evolution of the staggered magnetization
\begin{eqnarray}
m_{\rm s}=\frac{1}{N}\sum^N_{j=1}(-1)^jS^z_j,
\end{eqnarray}
which is a measure of the antiferromagnetic order within the XXX bath.
\begin{figure}
\includegraphics[width=.53\textwidth]{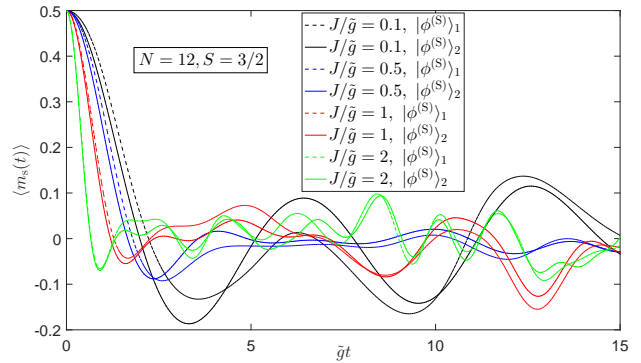}
\caption{Dynamics of the staggered magnetization $\langle m_{\rm s}(t)\rangle$ in a spin-$3/2$ Heisenberg star with $N=12$ bath spins. Two types of initial states for the central spin is used, i.e., $|\phi^{(\rm S)}\rangle_1=|S\rangle$ and $|\phi^{(\rm S)}\rangle_2=\frac{1}{\sqrt{2S+1}}(|S\rangle+|S-1\rangle+\cdots+|-S\rangle)$. The bath is initially prepared in the N\'eel state $|\mathrm{AF}\rangle=|\downarrow\uparrow\cdots\downarrow\uparrow\rangle$. }
\label{Fig3}
\end{figure}
Figure~\ref{Fig3} shows the staggered magnetization dynamics $\langle m_{\rm s}(t)\rangle$ for an XXX bath with $N=12$ sites and for a central spin of size $S=3/2$. For both types of the central-spin initial states $|\phi^{(\rm S)}\rangle_1$ and $|\phi^{(\rm S)}\rangle_2$, we see that $\langle m_{\rm s}(t)\rangle$ decays more rapidly as the intrabath coupling $J$ increases, which is consistent with the case of inhomogeneous system-bath couplings. Qualitatively, it is the nearest-neighbor intrabath coupling that mainly controls the short-time dynamics of the staggered magnetization.
\begin{figure}
\includegraphics[width=.53\textwidth]{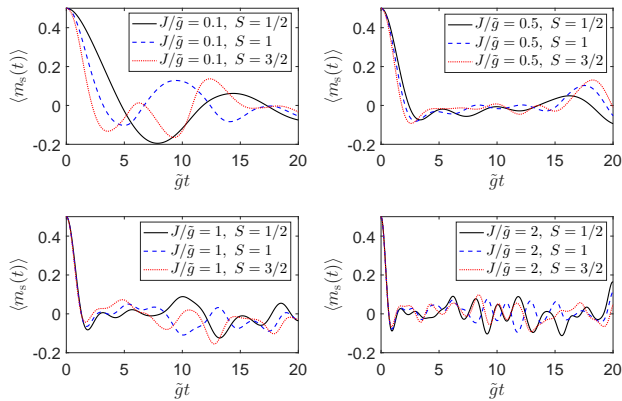}
\caption{Dynamics of the staggered magnetization $\langle m_s(t)\rangle$ in a spin-$S$ Heisenberg star with $N=12$ bath spins. Results for $S=1/2$, $1$, and $3/2$ are shown for fixed $J/\tilde{g}$. The initial state of the central spin is chosen as $|\phi^{(\rm S)}\rangle_1=|S\rangle$  and the bath is initially prepared in the N\'eel state $|\mathrm{AF}\rangle=|\downarrow\uparrow\cdots\downarrow\uparrow\rangle$.  }
\label{Fig4}
\end{figure}
For a fixed $J/\tilde{g}$, we find that the state $|\phi^{(\rm S)}\rangle_2$ induces a faster initial decay of $\langle m_{\rm s}(t)\rangle$ compared to $|\phi^{(\rm S)}\rangle_1$ since the former is more widely distributed in the Hilbert space. However, we observe that the difference between the two becomes smaller and smaller as we enter the strong intrabath coupling regime, where the system-bath coupling can be viewed as a perturbation, making the dynamics insensitive to the initial state of the central spin.
\par In Fig.~\ref{Fig4} we plot $\langle m_{\rm s}(t)\rangle$ for different values of $S$ and intrabath coupling $J$. Generally, a larger $S$ induces a faster initial decay of $\langle m_{\rm s}(t)\rangle$ since there are $2S+1$ channels for the central spin to interact with the XXX bath. This behavior is similar to that obtained for inhomogeneous system-bath couplings~\cite{Wu2022}. As expected, the deviation in the dynamics for different $S$'s becomes smaller when the intrabath coupling is large enough (lower panels of Fig.~\ref{Fig4}).
\subsection{Central spin dynamics in a modified Heisenberg star}
\par The polarization dynamics of a qubit coupled to a noninteracting spin bath prepared in the spin coherent state has been studied in several previous works~\cite{Dooley2013,Guan2019,PRA2020}. However, the case of a larger central spin coupled to an interacting spin bath is less studied. In this subsection, we will study the polarization dynamics of the central spin when the bath is prepared in a spin coherent state. As mentioned above, to get nontrivial intrabath coupling-induced central spin dynamics, we have to slightly modify the Heisenberg star given by Eq.~(\ref{Hstar}):
\begin{eqnarray}
\tilde{H}&=&\omega S_z+\sum^N_{j=1}[J(S^x_jS^x_{j+1}+S^y_jS^y_{j+1})+J'S^z_jS^z_{j+1}]\nonumber\\
&&+2g\sum^N_{j=1}\vec{S}\cdot\vec{S}_j,
\end{eqnarray}
where $\omega$ is an external magnetic field, $J$ and $J'$ are the in-plane and Ising parts of the intrabath coupling strength, respectively. Note that the bath angular momentum $\vec{L}^2$ is no longer conserved for $J\neq J'$. For $J=J'$ and $S=1/2$, $\tilde{H}$ is reduced to the model studied in Ref.~\cite{Yang2020}, which conserves $\vec{L}^2$. If one further sets $J=0$, then $\tilde{H}$ is reduced to a qubit$-$big-spin model~\cite{Guan2019}, whose dynamics can be analytically solved by using either a recurrence method~\cite{Guan2019,JSM2018} or an interaction-picture method~\cite{PRA2020}.
\par The initial state of the whole system reads
\begin{eqnarray}\label{psi0SCS}
|\psi(0)\rangle&=&|S\rangle\otimes|\hat{\Omega}\rangle,
\end{eqnarray}
where $|\hat{\Omega}\rangle$ is the spin coherent state of the bath defined by~\cite{PRA1972}
\begin{eqnarray}\label{Omega}
|\hat{\Omega}\rangle&=&\mathrm{e}^{-\mathrm{i}L_z\phi}\mathrm{e}^{-\mathrm{i}L_y\theta}|\frac{N}{2},\frac{N}{2}\rangle\nonumber\\
&=&\sum^N_{n=0}Q_n|\frac{N}{2},n-\frac{N}{2}\rangle,
\end{eqnarray}
with $Q_n=\frac{z^n}{(1+|z|^2)^{N/2}}\sqrt{C^n_N}$ and $z=\cot\frac{\theta}{2}\mathrm{e}^{-\mathrm{i}\phi}$. Here, $|\frac{N}{2},n-\frac{N}{2}\rangle$ is the Dicke state belonging to $l=N/2$ and has magnetization $l_m=n-N/2$.
\begin{figure}
\includegraphics[width=.49\textwidth]{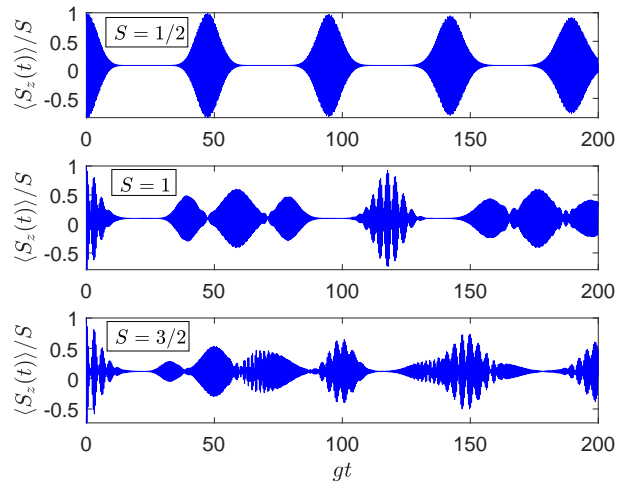}
\caption{Dynamics of the central-spin polarization $\langle S_z(t)\rangle/S$ for $J=J'$ and an XXX bath prepared in the spin coherent state $|\hat{\Omega}\rangle$. Parameters: $N=14$, $\theta=\frac{\pi}{2}$, $\phi=0$, and $\omega=g$.}
\label{Fig5}
\end{figure}
\par Let us first look at the case of an XXX bath with $J=J'$. We have demonstrated that in this case the central spin dynamics is independent of the value of $J$, which can be seen more straightforwardly by noting that the spin coherent state $|\hat{\Omega}\rangle$ is an eigenstate of $JH_{\rm b}$ with eigenvalue $NJ/4$:
\begin{eqnarray}
&&JH_{\rm b}|\hat{\Omega}\rangle=\mathrm{e}^{-\mathrm{i}L_z\phi}\mathrm{e}^{-\mathrm{i}L_y\theta}JH_{\rm b}|\frac{N}{2},\frac{N}{2}\rangle=\frac{NJ}{4}|\hat{\Omega}\rangle.\nonumber
\end{eqnarray}
It is thus necessary to go beyond the isotropic point $J=J'$ in order to observe nontrivial polarization dynamics induced by the intrabath coupling. Nevertheless, let us first study the effect of the value of $S$ on the central spin polarization dynamics for $J=J'$.
\par The top panel of Fig.~\ref{Fig5} shows the polarization dynamics $\langle S_z(t)\rangle/S$ of an $S=1/2$ central spin for $J=J'$ and under the resonant condition $\omega=g$~\cite{Guan2019}. It can be seen that the polarization exhibits the so-called collapse-revival behavior and the revival peaks occur at $gt\approx m N\pi ~(m\in \mathbb Z )$, recovering the analytical results presented in Ref.~\cite{Guan2019}. The middle and bottom panels of Fig.~\ref{Fig5} show $\langle S_z(t)\rangle/S$ for $S=1$ and $S=3/2$, respectively. The polarization still shows collapses and revivals during the evolution, but with rich fine structures. For example, the initial revival region seems show $2S$ discrete sub-peaks before the first collapse occurs. These structures reappear after the regular revival region consisting of $2S+1$ packets. We note that similar polarization dynamics is observed in Ref.~\cite{JSM2018} for a spin-1 central spin homogeneously coupled to a noninteracting spin bath.
\begin{figure}
\includegraphics[width=.52\textwidth]{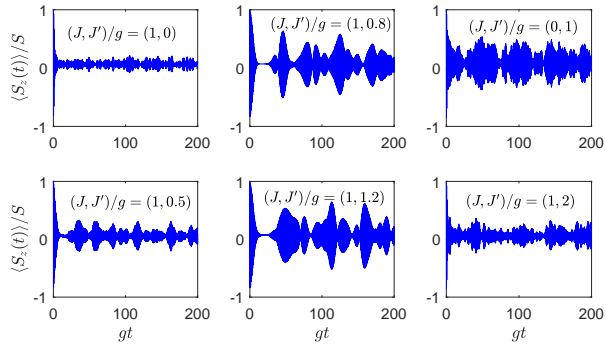}
\caption{Polarization dynamics of a qubit homogeneously coupled to an XXZ bath with $J\neq J'$. Parameters: $N=14$, $\theta=\frac{\pi}{2}$, $\phi=0$, and $\omega=g$.}
\label{Fig6}
\end{figure}
\par To see the effect of the anisotropy of the intrabath coupling on the polarization dynamics, we plot in Fig.~\ref{Fig6} $\langle S_z(t)\rangle/S$ for $S=1/2$ and several pairs of $(J/g,J'/g)$. It can be seen that the collapse-revival behaviors are generally destroyed, although for $(J,J')/g=(1,0.8)$ and $(1,1.2)$ there is some evidence of collapse (middle column of Fig.~\ref{Fig6}) at short time since they are close to the isotropic point $J'/J=1$. If we separate the term $(J'-J)\sum^N_{j=1}S^z_jS^z_{j+1}$ out of $\tilde{H}$, it is easy to check that this term does not commute with the remaining part of $\tilde{H}$. As a result, the dynamics depends not only on $J'-J$ but also on $J$ (right column of Fig.~\ref{Fig6}). Actually, since the term $(J'-J)\sum^N_{j=1}S^z_jS^z_{j+1}$ breaks the conservation of $\vec{L}^2$, the time-evolved state will run out of the $l=N/2$ subspace, making the collapse-revival phenomena fragile with respect to anisotropic intrabath coupling.
\section{Conclusions}\label{SecVI}
\label{sec-final}
\par In this work, we generalize the spin-1/2 Heisenberg proposed by Richter and Voigt~\cite{RV} to the case of arbitrary $S<N/2$. Compared with the spin-1/2 counterpart, both the ground-state  and the dynamical behaviors are found to have richer structures. In principle, the eigenenergies and eigenstates of the system can be obtained by block diagonalizing the Hamiltonian using the Bethe ansatz solution of the XXX bath, yielding invariant subspaces whose dimensions do not exceed $2S+1$. Based on the four conserved quantities of the model, we obtain all the eigenenergies of the model. The expressions of these eigenenergies differ depending whether $S$ is larger or smaller than the bath angular momentum $l$. The sub-ground state energies for fixed $l$ depend only on the quantum number $l$. The evolutions of the ground-state energy and the associated bath angular momentum are numerically analyzed when the intrabath coupling $J$ and the central-spin size $S$ are varied. We explain the observed behaviors of these quantities in the weak and strong intrabath coupling limits. We also derive closed-form expressions for the degenerate sub-ground states in each $l$-subspace.
\par We then study the real-time dynamics of the spin-$S$ Heisenberg star. Since the bath Hamiltonian commutes with the whole Hamiltonian, the intrabath coupling has no effect on the central spin dynamics if the system is prepared in a pure state. We thus turn to study the antiferromagnetic order dynamics within the XXX bath. Following Ref.~\cite{Wu2022}, we set the bath initial state to be a N\'eel state and investigate how the staggered magnetization evolves under the combined influence of the intrabath coupling and the system-bath coupling. We study the effects of the central spin initial state, the central spin size, and the system-bath coupling strength on the staggered magnetization dynamics and find similar behaviors to the inhomogeneous coupling case~\cite{Wu2022}.
\par  We finally study the central-spin polarization for a bath prepared in a spin coherent state. This is motivated by several recent works in which the polarization dynamics of a spin-1/2 coupled to a noninteracting spin bath is thoroughly studied~\cite{Dooley2013,Guan2019,PRA2020}. To observe nontrivial polarization dynamics that depends on the intrabath coupling, we extend the spin-$S$ Heisenberg star by including a Zeeman term of the central spin and the anisotropy in the intrabath coupling. At the isotropic point of the bath, we find that the polarization dynamics for $S>1/2$ exhibits collapse-revival behaviors with fine structures. However, for a spin bath with anisotropic coupling, the collapse-revival phenomena is generally found to be destroyed.
\par As an exactly soluble model, there are some other aspects of the spin-$S$ Heisenberg star deserve further investigation. For example, it would be interesting to study the dynamics of entanglement and quantum Fisher information and to understand quantum metrology in the present model. The analytical calculation of spin correlations in the weak intrabath coupling limit should be appealing. These studies will be left for future works.

\noindent{\bf Acknowledgements:}
 This work was supported by the National Key R\&D Program of China under Grant No. 2021YFA1400803 and by the Natural Science Foundation of China (NSFC) under Grant No. 11705007.
 
\appendix
\begin{widetext}
\textbf{Appendix}
\par To derive Eq.~(\ref{Asmfinal}), we apply the Hamiltonian $H=JH_{\rm b}+g(\frac{1}{2}S_+L_-+\frac{1}{2}S_-L_++S_zL_z)$ to the eigenstate $|\psi_{E^{(\mathrm{gs})}(l),l-S,m,l}\rangle=\sum^S_{S_m=-S}A_{S_m}|S_m\rangle|\phi_{E^{(1)}_{\rm b},l,m-S_m}\rangle$:
\begin{eqnarray}
&&H|\psi_{E^{(\mathrm{gs})}(l),l-S,m,l}\rangle\nonumber\\
&=&\left[JH_{\rm b}+g\left(\frac{1}{2}S_+L_-+\frac{1}{2}S_-L_++S_zL_z\right)\right]\sum^S_{S_m=-S}A_{S_m}|S_m\rangle|\phi_{E^{(1)}_{\rm b},l,m-S_m}\rangle\nonumber\\
&=&JE^{(1)}_{\rm b}\sum^S_{S_m=-S}A_{S_m}|S_m\rangle|\phi_{E^{(1)}_{\rm b},l,m-S_m}\rangle+g\sum^S_{S_m=-S}S_m(m-S_m)A_{S_m}|S_m\rangle|\phi_{E^{(1)}_{\rm b},l,m-S_m}\rangle\nonumber\\
&&+\frac{g}{2}\sum^S_{S_m=-S}A_{S_m}\sqrt{(S-S_m)(S+S_m+1)(l+m-S_m)(l-m+S_m+1)}|S_m+1\rangle|\phi_{E^{(1)}_{\rm b},l,m-S_m-1}\rangle\nonumber\\
&&+\frac{g}{2}\sum^S_{S_m=-S}A_{S_m}\sqrt{(S+S_m)(S-S_m+1)(l-m+S_m)(l+m-S_m+1)}|S_m-1\rangle|\phi_{E^{(1)}_{\rm b},l,m-S_m+1}\rangle\nonumber\\
&=&JE^{(1)}_b\sum^S_{S_m=-S}A_{S_m}|S_m\rangle|\phi_{E^{(1)}_{\rm b},l,m-S_m}\rangle+g\sum^S_{S_m=-S}S_m(m-S_m)A_{S_m}|S_m\rangle|\phi_{E^{(1)}_{\rm b},l,m-S_m}\rangle\nonumber\\
&&+\frac{g}{2}\sum^{S}_{S_m=-S+1}A_{S_m-1}\sqrt{(S-S_m+1)(S+S_m)(l+m-S_m+1)(l-m+S_m)}|S_m\rangle|\phi_{E^{(1)}_{\rm b},l,m-S_m}\rangle\nonumber\\
&&+\frac{g}{2}\sum^{S-1}_{S_m=-S}A_{S_m+1}\sqrt{(S+S_m+1)(S-S_m)(l-m+S_m+1)(l+m-S_m)}|S_m\rangle|\phi_{E^{(1)}_{\rm b},l,m-S_m}\rangle\nonumber\\
&=&E^{(\rm gs)}(l)\sum^S_{S_m=-S}A_{S_m}|S_m\rangle|\phi_{E^{(1)}_{\rm b},l,m-S_m}\rangle.
\end{eqnarray}
By comparing the coefficients on both sides, we get
\par (1) For $S_m=S$,
\begin{eqnarray}
&&A_S[JE^{(1)}_{\rm b}+gS(m-S)-E^{(\rm gs)}(l)] +\frac{g}{2} A_{S-1}\sqrt{2S(l+m-S+1)(l-m+S)}= 0.
\end{eqnarray}
(2) For $-S<S_m<S$,
\begin{eqnarray}
&&A_{S_m}[JE^{(1)}_{\rm b}+gS_m(m-S_m)-E^{(\rm gs)}(l)] \nonumber\\
&&+\frac{g}{2}A_{S_m-1}\sqrt{(S-S_m+1)(S+S_m)(l+m-S_m+1)(l-m+S_m)}\nonumber\\
&&+\frac{g}{2}A_{S_m+1}\sqrt{(S+S_m+1)(S-S_m)(l-m+S_m+1)(l+m-S_m)}=0.
\end{eqnarray}
(3) For $S_m=-S$,
\begin{eqnarray}
&&A_{-S}[JE^{(1)}_{\rm b}-gS(m+S)-E^{(\rm gs)}(l)] +\frac{g}{2} A_{-S+1}\sqrt{2S(l-m-S+1)(l+m+S)}=0.
\end{eqnarray}
\par By using $E^{(\rm gs)}(l)=JE^{(1)}_{\rm b}-gS(l+1)$, we have\\
(1) $S_m=S$:
\begin{eqnarray}\label{SmS}
&&A_S\sqrt{2 S(l+m-S+1)} +  A_{S-1}\sqrt{l-m+S}= 0.
\end{eqnarray}
(2) $-S<S_m<S$:
\begin{eqnarray}\label{SSmS}
&&2A_{S_m}[ S_m(m-S_m)+ S(l+1)] \nonumber\\
&&+ A_{S_m-1}\sqrt{(S-S_m+1)(S+S_m)(l+m-S_m+1)(l-m+S_m)}\nonumber\\
&&+ A_{S_m+1}\sqrt{(S+S_m+1)(S-S_m)(l-m+S_m+1)(l+m-S_m)}=0.
\end{eqnarray}
(3) $S_m=-S$:
\begin{eqnarray}\label{SmmS}
&&A_{-S}\sqrt{2 S(l-m-S+1)} +  A_{-S+1}\sqrt{ l+m+S }=0.
\end{eqnarray}
To solve these coupled system of equations, we note that Eq.~(\ref{SmS}) and (\ref{SmmS}) give (note that $l-m-S\geq0$ and $l+m-S\geq 0$)
\begin{eqnarray} \label{ASmAS}
\frac{A_{S-1}}{A_S}=-\sqrt{\frac{2S(l+m-S+1)}{l-m+S}},~~\frac{A_{-S}}{A_{-S+1}}=-\sqrt{\frac{l+m+S}{2S(l-m-S+1)}}.
\end{eqnarray}
Setting $S_m=S-1$ in Eq.~(\ref{SSmS}) gives
\begin{eqnarray}\label{Smm1}
&&2 [ (S-1)(m-S+1)+ S(l+1)]  + \frac{A_{S-2}}{A_{S-1}}\sqrt{2(2S-1)(l+m-S+2)(l-m+S-1)}\nonumber\\
&& + \frac{A_{S}}{A_{S-1}}\sqrt{2S (l-m+S)(l+m-S+1)}=0.
\end{eqnarray}
Combining Eqs.~(\ref{ASmAS}) with (\ref{Smm1}) gives
\begin{eqnarray} \label{AS2AS10}
  \frac{A_{S-2}}{A_{S-1}}  =-\sqrt{\frac{(2S-1)(l+m-S+2)}{2(l-m+S-1)}}.
\end{eqnarray}
The forms of Eqs.~(\ref{ASmAS}) and (\ref{AS2AS10}) suggest the following ansatz:
\begin{eqnarray} \label{AS2AS1}
  \frac{A_{S_m}}{A_{S_m+1}}  =-\sqrt{\frac{(S+S_m+1)(l+m-S_m)}{(S-S_m)(l-m+S_m+1)}}.
\end{eqnarray}
It is straightforward to verify that the above ansatz indeed solves Eqs.~(\ref{SSmS}) for all $-S<S_m<S$.
\par Starting with $A_S$, we find after iteration
\begin{eqnarray} \label{AsmApp}
A_{S_m}=(-1)^{S-S_m}\sqrt{C^{S+S_m}_{2S}}\sqrt{\frac{(l+m-S_m)!(l-m+S_m)!}{(l+m-S)!(l-m+S)!}}A_S.
\end{eqnarray}
By inserting Eq.~(\ref{AsmApp}) into the wave function, we obtain the unnormalized sub-ground state given by Eq.~(\ref{Asmfinal}) in the main text.
\par If we choose $m=l-S$, then
\begin{eqnarray}
&&|\psi_{E^{(\mathrm{gs})}(l),l-S,l-S,l}\rangle\nonumber\\
&=&\sum^S_{S_m=-S}(-1)^{S-S_m}  \sqrt{\frac{(2S)!}{(S+S_m)!(S-S_m)!}\frac{(2l-S-S_m)!(S+S_m)!}{(2l-2S)!(2S)!}}|S_m\rangle|\phi_{E^{(1)}_{\rm b},l,l-S-S_m}\rangle\nonumber\\
&=&\sqrt{\frac{1}{(2l-2S)!}}\sum^S_{S_m=-S}(-1)^{S-S_m}  \sqrt{\frac{(2l-S-S_m)! }{ (S-S_m)! }}|S_m\rangle|\phi_{E^{(1)}_{\rm b},l,l-S-S_m}\rangle,
\end{eqnarray}
whose squared norm is
\begin{eqnarray}
&&\langle\psi_{E^{(\mathrm{gs})}(l),l-S,l-S,l} |\psi_{E^{(\mathrm{gs})}(l),l-S,l-S,l}\rangle\nonumber\\
&=&  \frac{1}{(2l-2S)!}\sum^S_{S_m=-S}  \frac{(2l-S-S_m)! }{ (S-S_m)! }\nonumber\\
&=& \frac{1}{(2l-2S)!}\sum^{2S}_{S'_m=0}  \frac{(2l-2S+S'_m)! }{ S'_m! }\nonumber\\
&=&   \frac{1}{(2l-2S)!}\frac{(2l+1)!(2S+1)}{(2l-2S+1)(2S+1)!}\nonumber\\
&=&   \frac{(2l+1)! }{(2l-2S+1)!(2S )!}\nonumber\\
&=& C^{2S}_{2l+1}.
\end{eqnarray}
Thus, the normalized state $|\psi_{E^{(\mathrm{gs})}(l),l-S,l-S,l}\rangle$ is given by Eq.~(\ref{AsmfinalHW}).
\end{widetext}

\end{document}